\documentclass{aa}  
\usepackage{graphicx}
\usepackage{changes}
\usepackage{txfonts}
\usepackage{lscape}
\usepackage{natbib}
\newcommand{\figureloc}{./}

\begin{document}
\title{High-precision chemical abundances of Galactic building blocks. II.}
\subtitle{Revisiting the chemical distinctness of the Helmi streams}
\author{Tadafumi Matsuno \inst{1} 
\and 
Emma Dodd \inst{1}
\and
Helmer H. Koppelman \inst{2}
\and 
Amina Helmi \inst{1}
\and
Miho N. Ishigaki \inst{3}
\and
Wako Aoki \inst{3}
\and
Jingkun Zhao \inst{4}
\and
Zhen Yuan \inst{5}
\and
Kohei Hattori \inst{3,6}
}

\institute{
   Kapteyn Astronomical Institute, University of Groningen, Landleven 12, 9747 AD Groningen, The Netherlands\\
   \email{matsuno@astro.rug.nl}
     \and
   School of Natural Sciences, Institute for Advanced Study, 1 Einstein Drive, Princeton, NJ 08540, USA
  \and
   National Astronomical Observatory of Japan, 2-21-1 Osawa, Mitaka, Tokyo 181-8588, Japan 
   \and
   Key Lab of Optical Astronomy, National Astronomical
  Observatories, Chinese Academy of Sciences, A20 Datun Road, Chaoyang, Beijing 100101, China
  \and
 Universit\'{e} de Strasbourg, CNRS, Observatoire Astronomique de Strasbourg, UMR 7550, F-67000 Strasbourg, France
  \and
 Institute of Statistical Mathematics, 10-3 Midoricho, Tachikawa, Tokyo 190-0014, Japan
  }

\abstract
  {
  The Helmi streams are a kinematic substructure whose progenitor is likely a dwarf galaxy.
 Although 20 years have passed since their discovery, it is still unclear whether their members are chemically distinguishable from other halo stars in the Milky Way.
  } 
  {
  We aim to precisely characterize the chemical properties of the Helmi streams.
  } 
  {
  We analyzed high-resolution, high signal-to-noise ratio spectra for 11 Helmi stream stars through a line-by-line abundance analysis.
  We compared the derived abundances to homogenized literature abundances of the other halo stars, including those belonging to other kinematic substructures, such as Gaia-Enceladus and Sequoia.
  } 
  {
  Compared to typical halo stars, the Helmi stream members clearly show low values of [{X}/{Fe}] in elements produced by massive stars, such as Na and $\alpha$-elements.
  This tendency is seen down to metallicities of at least $[\mathrm{Fe/H}]\sim -2.2$, suggesting type~Ia supernovae already started to contribute to the chemical evolution at this metallicity.
  We find that the [{$\alpha$}/{Fe}] ratio does not evolve significantly with metallicity, making the Helmi stream stars less distinguishable from Gaia-Enceladus stars at $[\mathrm{Fe/H}]\gtrsim -1.5$.
  The almost constant but low value of [{$\alpha$}/{Fe}] might be indicative of quiescent star formation with low efficiency at the beginning and bursty star formation at later times.
  We also find extremely low values of [{Y}/{Fe}] at low metallicity, providing further support for the claim that light neutron-capture elements are deficient in Helmi streams.
  While Zn is deficient at low metallicity, it shows a large spread at high metallicity.
  The origin of the extremely low Y abundances and Zn variations remains unclear.
  } 
  {
  The Helmi stream stars are distinguishable from the majority of the halo stars if homogeneously derived abundances are compared.
  } 
\maketitle

\defcitealias{Matsuno2022a}{Paper I}
\defcitealias{Nissen2010}{NS10}
\defcitealias{Reggiani2017a}{R17}

\section{Introduction\label{sec:intro}}

\begin{table*}
  \caption{Summary of the data \label{tab:obs}}
  \centering
  \begin{tabular}{lrrrr}
\hline\hline
Object & Gaia EDR3 source id & $S/N_{4500\,\mathrm{\AA}}$ & $S/N_{5533\,\mathrm{\AA}}$ & $S/N_{6370\,\mathrm{\AA}}$ \\\hline
2447\_5952       &  2447968154259005952 & 80   & 108  & 173           \\
4998\_5552       &  4998741805354135552 & 72   & 83   & 78            \\
6170\_9904       &  6170808423037019904 & 87   & 88   & 77            \\
6615\_9776       &  6615661065172699776 & 97   & 161  & 122           \\
6914\_3008       &  6914409197757803008 & 73   & 86   & 120           \\
J1306+4154      &  1527475951701753984 & 50   & 65   & 53            \\
J1436+0929      &  1176187720407158912 & 134  & 110  & 91            \\
J1553+3909      &  1376687518318241536 & 84   & 73   & 55            \\
J1730+5309      &  1416077522383596160 & 84   & 63   & 42            \\
J1642+2041      &  4564066449004092928 & 63   & 60   & 86            \\
LP894-3         &  2891152566675457280 & 97   & 150  & 143           \\
\hline
  \end{tabular}
\tablefoot{We obtained new high-resolution spectra for all objects except LP894-3. The signal-to-noise ratios are converted to per $0.01\,\mathrm{\AA}$.}
 \end{table*} 

\begin{table*}
  \caption{Property of the targets \label{tab:targets}}
  \centering
  \begin{tabular}{lrrrrrr}
\hline\hline
Object     & $\pi$ & $\sigma(\pi)$ & $G$    & $B_p-R_p$ & $E(B-V)$ & RV  \\
           & (mas) & (mas) & (mag)    & (mag) & (mag) & ($\mathrm{km\,s^{-1}}$)  \\\hline
2447\_5952  & 2.817 & 0.021           & 11.963 & 0.677     & 0.096    & 145.5  \\
4998\_5552  & 3.463 & 0.018           & 12.561 & 0.784     & 0.000\tablefootmark{a}    & -187.6  \\
6170\_9904  & 3.743 & 0.018           & 10.573 & 0.728     & 0.064    & -122.4  \\
6615\_9776  & 4.984 & 0.014           & 12.064 & 0.768     & 0.010    & -276.9  \\
6914\_3008  & 3.195 & 0.017           & 12.622 & 0.720     & 0.015    & 113.9  \\
J1306+4154 & 1.470 & 0.013           & 13.499 & 0.668     & 0.046    & -294.7  \\
J1436+0929 & 3.294 & 0.019           & 12.172 & 0.671     & 0.006    & -301.4  \\
J1553+3909 & 1.944 & 0.013           & 13.619 & 0.741     & 0.018    & -288.3  \\
J1642+2041 & 0.997 & 0.013           & 13.479 & 0.765     & 0.070    & -246.3  \\
J1730+5309 & 0.751 & 0.011           & 13.967 & 0.817     & 0.082    & 70.9  \\
LP894-3    & 5.347 & 0.038           & 11.090 & 0.710     & 0.010    & 303.0\tablefootmark{b}  \\
\hline
  \end{tabular}
\tablefoot{
\tablefoottext{a}{This object is not covered by \citet{Green2019a}. Since \citet{Schlegel1998a} provide $E(B-V)=0.01$ and since this object is nearby, we assumed $E(B-V)=0.0$ for this object.}
\tablefoottext{b}{From \citet{Ishigaki2012}.}
}
 \end{table*} 

The Helmi streams were discovered by \citet{Helmi1999a} as a kinematic substructure among halo stars, which is an expected signature of past galaxy accretion.
They were thus one of the first pieces of evidence for the hierarchical formation of the Milky Way expected from the standard cosmology.
The streams were discovered as an overdensity in the plane of angular momenta in a sample of $<100$ low-metallicity stars.
Later studies also confirmed their existence \citep{Chiba2000a,Kepley2007a,Myeong2018d,Koppelman2018a}.
Their progenitor was likely a dwarf galaxy with a stellar mass of $\sim 10^8\,\mathrm{M_{\odot}}$ and a mean metallicity of $[\mathrm{Fe/H}]\sim -1.5$ that was accreted to the Milky Way around $5-8\,\mathrm{Gyr}$ ago \citep{Kepley2007a,Koppelman2019b}.

Even though the streams were discovered more than 20 years ago, their chemical properties have not received much attention until recently. 
\citet{Roederer2010a} studied the chemical abundances of 12 stars in the stream, concluding that the stars have similar abundance ratios as the rest of the halo stars.
In recent years, \citet{Limberg2021a}, \citet{Gull2021a}, \citet{Aguado2021b}, and \citet{Nissen2021a} conducted chemical analysis for the Helmi stream members.
While \citet{Gull2021a} reach the same conclusion as \citet{Roederer2010a}, the other studies hint at possible chemical peculiarities of the streams. 
For example, the Sr abundance seems to be much lower than typical for halo stars at $[\mathrm{Fe/H}]\lesssim -2$ \citep{Aguado2021b}; two stars in a binary system in the Helmi streams, namely G112-43 and G112-44, have enhanced ratios of [{Mn}/{Fe}], [{Ni}/{Fe}], [{Cu}/{Fe}], and [{Zn}/{Fe}] compared to other accreted stars in the Milky Way \citep{Nissen2021a}; and the [$\alpha$/{Fe}] ratio seems to be decreasing at $[\mathrm{Fe/H}]\gtrsim -2$ \citep{Limberg2021a,Aguado2021b}.

Many of the previous comparisons made use of heterogeneous literature compilation.
They were therefore limited by systematic uncertainties, which can be as large as 0.4 dex in [{Fe}/{H}] and 0.2 dex in [{X}/{Fe}].
This can easily obscure or falsely create abundance peculiarities, as we will see later in this paper.
Another complication arises because the literature sample was often a mixture of accreted and in situ stars.
As \citet[][hereafter, NS10]{Nissen2010} showed from their homogeneously analyzed samples, the Milky Way halo contains two major populations, accreted and in situ stars, which differ in abundance ratios.
Without understanding the origin of the comparison sample, it remains unclear if the chemical peculiarities of the Helmi stream stars, if any, indicate simply that they are accreted or that the progenitor had a unique evolution.
\citet{Nissen2021a} is the only exception in the sense that they compared Helmi stream members with accreted stars using homogeneously derived chemical abundance.
However, larger samples could aid our understanding of the interesting abundance pattern reported in the two stars in a binary system.

Throughout this study, we aim to precisely characterize the chemical abundances of the Helmi stream stars and to compare them with homogeneous abundances of other components in the Milky Way halo, including Gaia-Enceladus, in situ stars (NS10; Reggiani et al. 2017, hereafter R17), and Sequoia \citep[][hereafter, Paper I]{Matsuno2022a}.
The precise chemical abundance ratios tell us about the progenitor's properties, but the chemical distinctness of kinematic substructures, if found, would also enable us to ``chemically tag'' stars originating from the same progenitor.
In \citetalias{Matsuno2022a}, we show that a combination of a line-by-line abundance analysis for a carefully selected sample and homogenization of literature abundances using standard stars is a powerful way to precisely characterize the chemical properties of kinematic substructures.

We studied the chemical abundance of the Helmi stream stars in the same way as in \citetalias{Matsuno2022a}. 
Namely, we analyzed high signal-to-noise ratio spectra for 11 stars in the Helmi streams, measured their chemical abundance through a line-by-line analysis with careful treatment of uncertainty, and compared them with other halo stars in \citetalias{Nissen2010} and \citetalias{Reggiani2017a} after homogenizing all the abundances.
We describe the observation, data reduction, targets, and stars in the literature in Section~\ref{sec:data}.
In Section~\ref{sec:abundance}, after briefly explaining our approach for the chemical abundance measurement and homogenization, we present chemical abundance ratios of the Helmi stream stars.
After providing discussions in Section~\ref{sec:discussion}, we summarize our findings in Section~\ref{sec:conclusion}.

\section{Data\label{sec:data}}

\subsection{Observation}
\begin{table*}
  \caption{Kinematics of the targets \label{tab:kinematics}}
  \centering
  \begin{tabular}{l*{9}{r}}
\hline\hline
Object     & $v_z$  & $\sigma(v_z)$ & $L_z$ & $\sigma(L_z)$ & $L_{\rm perp}$ & $\sigma(L_{\rm perp})$ & $E/10^5$    & $\sigma(E)/10^5$ & subclass\tablefootmark{a} \\
           & ($\mathrm{km\,s^{-1}}$)  & ($\mathrm{km\,s^{-1}}$) & ($\mathrm{kpc\,km\,s^{-1}}$) & ($\mathrm{kpc\,km\,s^{-1}}$) & ($\mathrm{kpc\,km\,s^{-1}}$) & ($\mathrm{kpc\,km\,s^{-1}}$) &  ($\mathrm{km^2\,s^{-2}}$)   & ($\mathrm{km^2\,s^{-2}}$) &                           \\\hline
2447\_5952  & -216.7 & 1.1           & 1278  & 11            & 1872           & 9                      & -1.022 & 0.007       & ... \\
4998\_5552  & 230.1  & 1.1           & 1020  & 6             & 1846           & 9                      & -1.429 & 0.002       & ... \\
6170\_9904  & -272.8 & 1.3           & 1200  & 9             & 2201           & 11                     & -1.342 & 0.003       & hiL\\
6615\_9776  & 213.8  & 0.9           & 1345  & 2             & 1757           & 8                      & -1.339 & 0.003       & loL\\
6914\_3008  & -283.9 & 1.5           & 1437  & 8             & 2279           & 12                     & -1.275 & 0.004       & hiL\\
J1306+4154 & -281.3 & 0.9           & 1288  & 2             & 2363           & 7                      & -1.293 & 0.002       & hiL\\
J1436+0929 & -249.4 & 0.9           & 1226  & 4             & 2050           & 7                      & -1.312 & 0.003       & hiL\\
J1553+3909 & -268.6 & 0.8           & 1304  & 6             & 2208           & 7                      & -1.299 & 0.002       & hiL\\
J1642+2041 & -217.0 & 1.2           & 1339  & 7             & 1737           & 9                      & -1.392 & 0.003       & loL\\
J1730+5309 & 238.3  & 2.4           & 1423  & 14            & 1965           & 19                     & -1.384 & 0.004       & loL\\
LP894-3    & -219.8 & 0.7           & 1015  & 6             & 1813           & 6                      & -1.283 & 0.003       & ...\\
\hline
  \end{tabular}
\tablefoot{
\tablefoottext{a}{Subclass defined by \citet{Dodd2022a}. In the last column, ''hiL'' indicates a clump with a higher $L_{\rm perp}$ and ''loL'' one with a lower $L_{\rm perp}$.}
}
 \end{table*} 

\begin{figure*}
\includegraphics[width=\textwidth]{\figureloc 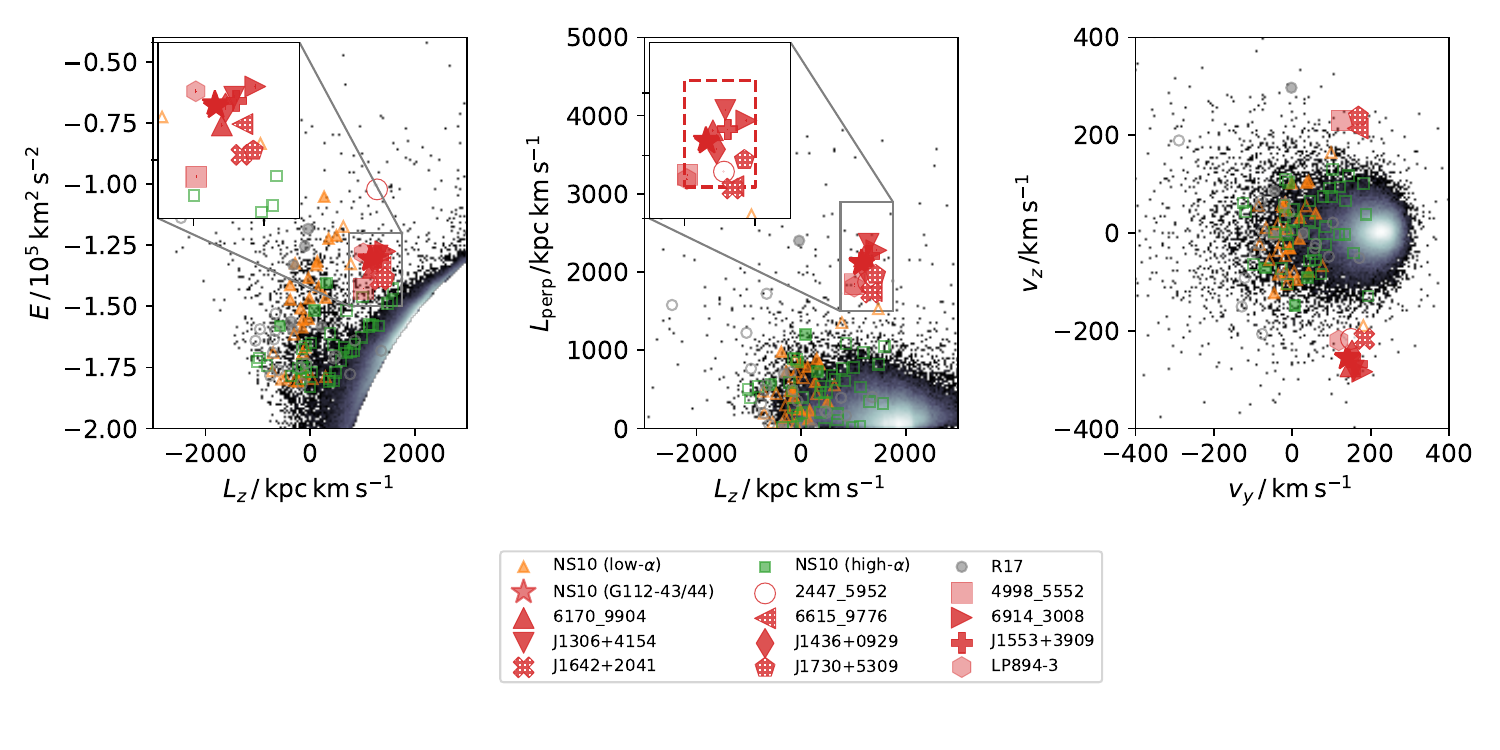}
\caption{Distribution of stars in kinematic spaces, namely $E-L_z$, $L_{\rm perp}-L_z$, and $v_y-v_z$. The insets in the left two panels show better  distribution of the stars that are part of the Helmi streams. 
The Helmi stream selection from \citet{Koppelman2019b} is shown as a dashed box in the inset of the middle panel. 
Stars are hatched according to the subgroups found by \citet{Dodd2022a}. Three stars (2447\_5952, 4998\_5552, and LP894-3) are not associated with either of the subgroups. An open symbol is assigned only to 2447\_5952 since it has much larger energy than the other Helmi stream stars.
We used filled symbols for stars from \citetalias{Nissen2010} and  \citetalias{Reggiani2017a} if they satisfy the kinematic selection for Gaia-Enceladus \citepalias[see][]{Matsuno2022a}. 
\label{fig:kinematics}}
\end{figure*}

\begin{figure}
\includegraphics[width=0.5\textwidth]{\figureloc 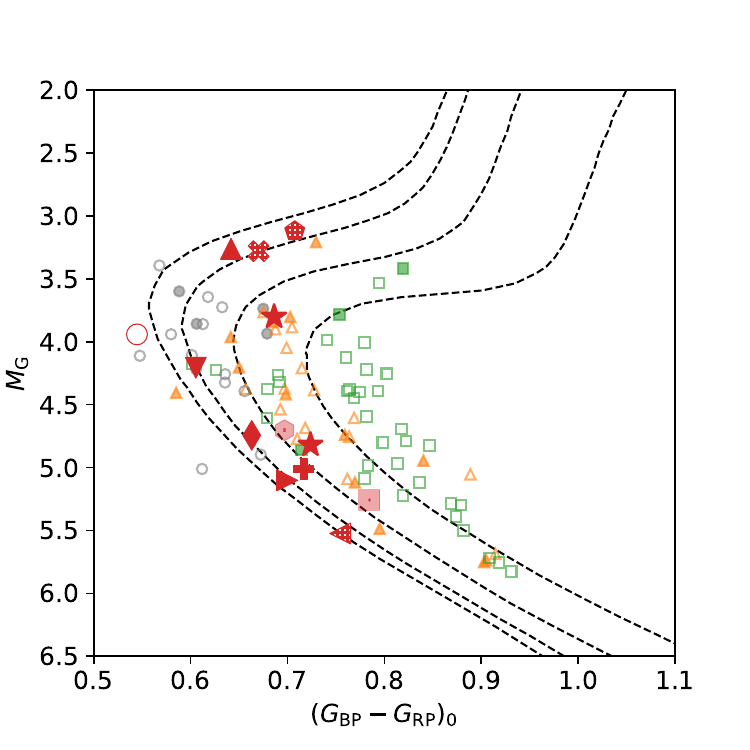}
\caption{Location of the stars in the color-magnitude diagram. The symbols follow those of Figure~\ref{fig:kinematics}. The four isochrones are from the PARSEC tracks with the age of 12 Gyr and $[\mathrm{Fe/H}]=-2.0,\,-1.5,\,-1.0,\,\mathrm{and}\,-0.5$.\label{fig:cmd}}
\end{figure}

We obtained high-resolution spectra for ten stars with the High Dispersion Spectrograph \citep{Noguchi2002} on the Subaru telescope\footnote{Gaia EDR3 1909569058536197760 was also observed but not analyzed since it is a double-lined spectroscopic binary.}.
The selection and properties of the targets are described in the following subsection.
The observations were conducted on August 9-10, 2020, with the standard setup of the HDS (StdYd), which yields a wavelength coverage from $\sim4000\,\mathrm{\AA}$ to $\sim 6800\,\mathrm{\AA}$.
We used the image slicer \#2 \citep{Tajitsu2012a}, which provides a resolving power of $R\sim 80,000$. 
The data reduction was performed using an \texttt{IRAF} script\footnote{IRAF is distributed by the National Optical Astronomy Observatory, which is operated by the Association of Universities for Research in Astronomy (AURA) under a cooperative agreement with the National Science Foundation}, \texttt{hdsql}\footnote{\url{http://www.subarutelescope.org/Observing/Instruments/HDS/hdsql-e.html}}, including CCD linearity correction, scattered light subtraction, aperture extraction, flat-fielding, wavelength calibration, and heliocentric velocity correction.
Information about the spectra is summarized in Table~\ref{tab:obs}.

We also searched archives for high signal-to-noise ratio and high-resolution spectra for stars that satisfy our selection criteria described in the following subsection. 
We find that LP894-3 satisfies our selection and has spectra in the Subaru telescope archive.
We, therefore, analyzed the archival spectrum of this star.
The spectrum used is the same as that used by \citet{Ishigaki2012}.

\subsection{Targets\label{sec:targets}}

Targets were selected based on the orbital angular momenta of the stars following \citet{Koppelman2019b}.
Namely, the selection box was defined with the component of the angular momentum along the $z$-axis of the Milky Way ($L_z$) and $L_{\rm perp}$, which is given by $\sqrt{L_x^2+L_y^2}$. 
We also restricted our sample to those around the turn-off region using Gaia DR2 photometry. 
The exact selection criteria were $1000<L_z/\mathrm{kpc\,km\,s^{-1}}<1500$ and $1750<L_{\rm perp}/\mathrm{kpc\,km\,s^{-1}}<2600$ \citep[the kinematic selection, corresponding to the box A of][]{Koppelman2019b}, and $0.5<G_{BP}-G_{RP}<0.95$ and $3.2<G_{\rm abs}<5.8$ (the color-magnitude selection), where $G_{BP}$, $G_{RP}$, and $G_{\rm abs}$ are Gaia $BP$ and $RP$ magnitude, and the absolute magnitude in the Gaia $G$ band, respectively. 

Initially, for the target selection, we used Gaia DR2 astrometry, photometry, and radial velocity \citep{GaiaCollaboration2018a}, and Large Sky Area Multi-Object Fiber Spectroscopic Telescope (LAMOST) survey DR5 radial velocity.
We assumed $U_\odot = -11.1\,\mathrm{km\,s^{-1}}$, $W_\odot=7.3\,\mathrm{km\,s^{-1}}$ \citep{Schoenrich2010a}, $R_0=8.21\,\mathrm{kpc}$ \citep{McMillan2017a}, and $z_0=20.8\,\mathrm{pc}$ \citep{Bennett2019a} for the solar velocity and position.
Adopting the proper motion measurement for Sgr A$^*$ \citep{Reid2004a}, we assumed the sun is moving at $245.3\,\mathrm{km\,s^{-1}}$ in the direction of the Galactic rotation.
When calculating the orbital energies of the stars, we adopted the Milky Way potential of \citet{McMillan2017a}.

Information about the kinematic and photometric properties of the targets is summarized in Tables~\ref{tab:targets} and \ref{tab:kinematics}, and visualized in Figures~\ref{fig:kinematics} and \ref{fig:cmd}.
We note that we updated the kinematic and photometric information with Gaia EDR3 photometry and astrometry \citep{GaiaCollaboration2021a}.
The radial velocities were also updated to the values we measured from our high-resolution spectra by comparing the observed wavelengths of neutral Fe lines with those from laboratory experiments.

Table~\ref{tab:kinematics} also includes information about subgroups among the Helmi stream stars.
We consider here two ways of subdividing the Helmi stream stars.
One is by the sign of $v_z$; \citet{Kepley2007a} and \citet{Koppelman2019b} used the asymmetry in the number of stars in each subgroup to estimate the accretion time of the progenitor of the Helmi streams.
Another subdivision follows that by \citet{Dodd2022a} in the space of $L_z$ and $L_{\rm perp}$.
They show that these two groups of stars with slightly different kinematics reflect the effect of an orbital resonance in the Milky Way potential.
In either definition, the two subgroups are considered parts of the same substructure, the Helmi streams.
In all figures, symbols are hatched according to the subdivision by \citet{Dodd2022a}.

\subsection{Literature}

We compared the chemical abundance of the Helmi stream stars with those in literature.
Since our approach described in Section~\ref{sec:abundance} is identical to that used in \citetalias{Matsuno2022a}, we can naturally compare our abundance in the present study with Gaia-Enceladus and in situ stars from \citetalias{Nissen2010} and \citetalias{Reggiani2017a} and with Sequoia stars from \citetalias{Matsuno2022a}.
Since abundances from these studies were homogenized using standard stars \citepalias[see ][]{Matsuno2022a}, precise comparisons are possible.
We refer to \citetalias{Matsuno2022a} for more details.

Among the stars studied by \citetalias{Nissen2010} and \citetalias{Reggiani2017a}, we find two stars that satisfy our kinematic selection criteria for the Helmi streams.
These two stars are G112-43 and G112-44 and were studied in detail by \citet{Nissen2021a} as part of the Helmi streams.
Although they are considered to form a binary system, their motions within the system do not affect the estimated orbital properties within the Milky Way since the separation between the two stars is large ($>2000\,\mathrm{au}$).

\section{Chemical abundances\label{sec:abundance}}
\begin{sidewaystable*}
  \caption{Stellar parameters\label{tab:parameters}}
  \centering
  \begin{tabular}{l*{14}{r}}
\hline\hline
Object & $T_{\rm eff}$ & $\sigma(T_{\rm eff})$ &  $\log g$ & $\sigma(\log g)$ & $v_t$ & $\sigma(v_t)$ &  $[\mathrm{Fe/H}]$ & $\sigma([\mathrm{Fe/H}])$ & $\rho_{T_{\rm eff},\log g}$ & $\rho_{T_{\rm eff},v_t}$ & $\rho_{T_{\rm eff},[\mathrm{Fe/H}]}$ &  $\rho_{\log g,v_t}$ & $\rho_{\log g,[\mathrm{Fe/H}]}$ & $\rho_{v_t,[\mathrm{Fe/H}]}$  \\
 & (K) & (K) &  (dex) & (dex) & ($\mathrm{km\,s^{-1}}$) & ($\mathrm{km\,s^{-1}}$) &  (dex) & (dex) &  &  & & & &  \\\hline
2447\_5952       & 6275 &   84 &  4.195&  0.040&  1.371&  0.101& -1.416&  0.028&  0.498&  0.547&  0.263&  0.058&  0.613& -0.337 \\
4998\_5552       & 5562 &   57 &  4.413&  0.058&  1.022&  0.154& -1.282&  0.040&  0.071&  0.695& -0.305& -0.268&  0.605& -0.579 \\
6170\_9904       & 5946 &   72 &  3.789&  0.043&  1.251&  0.102& -1.706&  0.033&  0.425&  0.605&  0.003&  0.138&  0.389& -0.395 \\
6615\_9776       & 5490 &   61 &  4.481&  0.033&  0.752&  0.226& -2.175&  0.026&  0.440&  0.836& -0.540&  0.239&  0.195& -0.782 \\
6914\_3008       & 5706 &   69 &  4.400&  0.034&  0.821&  0.157& -1.730&  0.025&  0.399&  0.836& -0.343&  0.128&  0.323& -0.500 \\
J1306+4154      & 6175 &  108 &  4.261&  0.051&  1.272&  0.114& -1.211&  0.029&  0.479&  0.696&  0.094&  0.180&  0.583& -0.221 \\
J1436+0929      & 6109 &   55 &  4.374&  0.031&  1.370&  0.104& -1.839&  0.038&  0.428&  0.634&  0.012&  0.166&  0.249& -0.244 \\
J1553+3909      & 5623 &   60 &  4.346&  0.036&  0.871&  0.184& -1.409&  0.022&  0.405&  0.745& -0.526&  0.175&  0.339& -0.749 \\
J1642+2041      & 5840 &   70 &  3.784&  0.089&  1.293&  0.083& -1.275&  0.043& -0.037&  0.762& -0.143& -0.446&  0.827& -0.501 \\
J1730+5309      & 5778 &   59 &  3.706&  0.071&  1.245&  0.067& -1.661&  0.039& -0.024&  0.587& -0.021& -0.541&  0.776& -0.502 \\
LP894-3         & 6035 &   83 &  4.341&  0.038&  1.321&  0.142& -1.505&  0.034&  0.561&  0.856&  0.154&  0.387&  0.370&  0.021 \\
\hline
\end{tabular}
\end{sidewaystable*}

\begin{table*}
  \caption{Linelist and line-by-line abundance\label{tab:linelist}}
  \centering
  \begin{tabular}{l*{7}{r}}\hline\hline
Object    & species & $\lambda$ & $\chi$ & $\log gf$ & $EW$  & $\sigma(EW)$ & $A(X)$\\
          &         &($\mathrm{\AA}$) & (eV) & (dex) & ($\mathrm{m\AA}$)  & ($\mathrm{m\AA}$) & (dex)\\\hline 
2447\_5952 & NaI     & 5682.633  & 2.102  & -0.706    & 5.1   & 0.5          & 4.675\\  
2447\_5952 & NaI     & 5889.959  & 0.000  & -0.193    & 172.0 & 8.0          & 4.558\\  
2447\_5952 & NaI     & 5895.910  & 0.000  & -0.575    & 151.6 & 7.0          & 4.505\\  
2447\_5952 & MgI     & 4167.271  & 4.346  & -0.746    & 82.4  & 3.8          & 6.385\\  
2447\_5952 & MgI     & 4730.040  & 4.340  & -2.379    & 5.3   & 0.6          & 6.472\\\hline\hline  
$\sigma(A)_{T_{\rm eff}}$ & $\sigma(A)_{\log g}$ & $\sigma(A)_{v_t}$ & $\sigma(A)_{[\mathrm{Fe/H}]}$ & $\sigma(A)_{EW}$ & $s_X$ & weight \\
(dex) & (dex) & (dex) & (dex) & (dex) & (dex) & \\\hline
0.028                     & 0.000                & 0.000             & 0.001                         & 0.049          & 0.000 & 285.037\\
0.084                     & -0.017               & -0.015            & 0.003                         & 0.098          & 0.000 & 22.804\\
0.084                     & -0.014               & -0.020            & 0.003                         & 0.103          & 0.000 & 22.571\\
0.041                     & -0.009               & -0.007            & 0.001                         & 0.061          & 0.000 & 124.161\\
0.030                     & 0.000                & 0.000             & 0.001                         & 0.055          & 0.000 & 170.647\\\hline
  \end{tabular}
\tablefoot{The full table is available online at the CDSr; only a portion of the table is shown here.}
 \end{table*}

\subsection{Abundance analysis\label{sec:analysis}}
\begin{figure*}
\includegraphics[width=\textwidth]{\figureloc 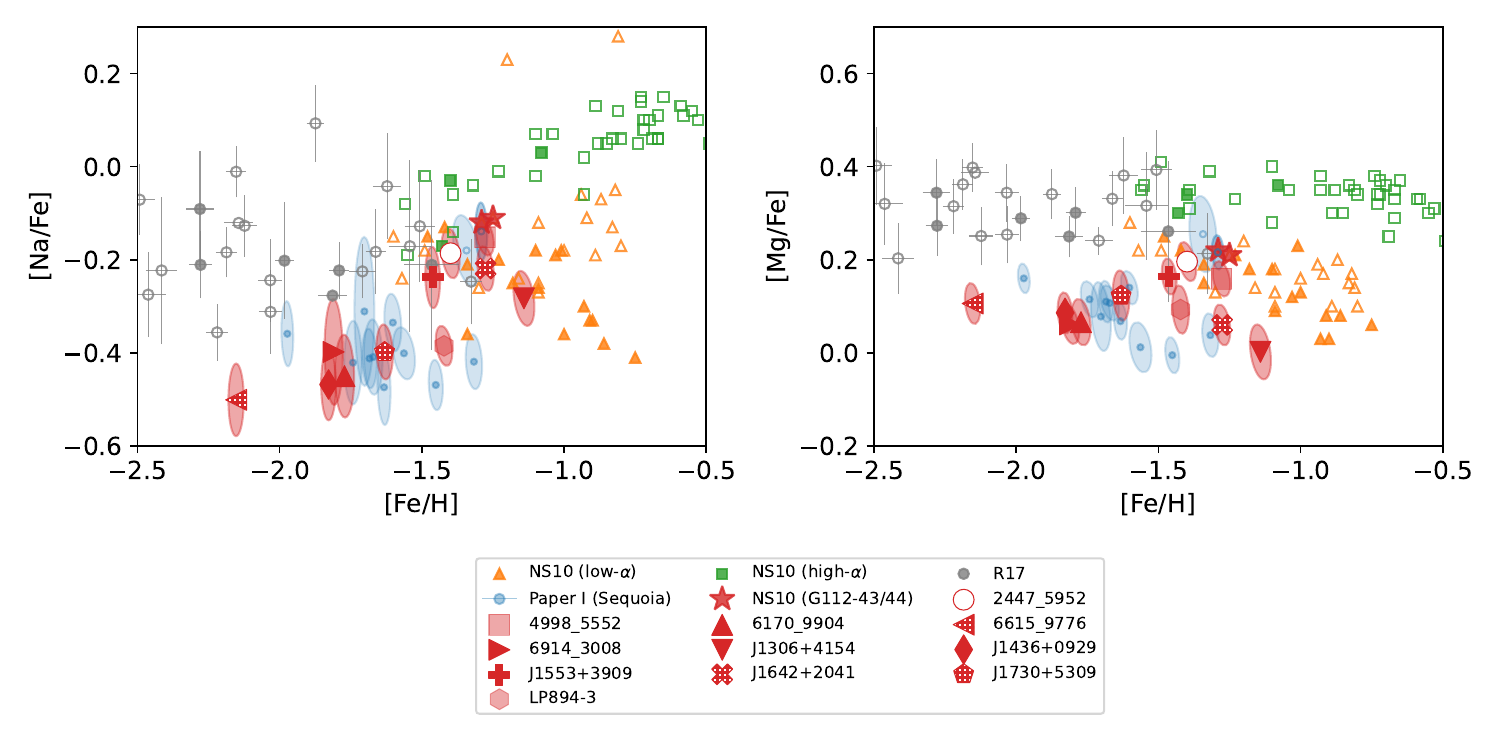}
\caption{[{Na}/{Fe}] and [{Mg}/{Fe}] abundance ratios of the Helmi stream stars (red symbols). Also plotted are halo stars from \citetalias{Nissen2010} and \citetalias{Reggiani2017a}, with Gaia-Enceladus stars shown with filled symbols. G112-43 and G112-44 from \citetalias{Nissen2010} are shown with red stars since their kinematics are consistent with the Helmi streams. We also show kinematically selected Sequoia stars from \citetalias{Matsuno2022a}.\label{fig:NaMg}}
\end{figure*}

We provide here a brief description of our analysis, and readers can refer to \citetalias{Matsuno2022a} for a complete and detailed explanation.
The abundance measurements were based on equivalent width measurements, where we assumed a Voigt profile for the line shape.
We applied spectral synthesis for Si, Mn, Zn, and Y.
Hyperfine structure splitting was considered for Na, Mn, and Ba.
Abundances were estimated based on the assumption of local thermodynamic equilibrium (LTE) using MOOG \citep{Sneden1973} and based on one-dimensional plane-parallel MARCS atmosphere models \citep{Gustafsson2008a}.
We applied non-LTE corrections of \citet{Lind2011a} to the obtained Na abundance through the INSPECT database\footnote{
Data obtained from the INSPECT database, version 1.0 (\url{www.inspect-stars.com}).}
\footnote{The Na abundances in the comparison literature are in LTE (NS10) and in non-LTE (R17). 
This inconsistency does not affect our conclusions since we put all the abundances onto the same scale and since NS10 used Na lines that require small non-LTE corrections ($\lesssim 0.1$ dex).}.

The effective temperature ($T_{\rm eff}$) was determined by minimizing the correlation between the excitation potentials and line-by-line abundances of neutral iron lines.
The microturbulent velocity ($v_t$) was determined similarly by minimizing the correlation between reduced equivalent widths and abundances.
The surface gravity ($\log g$) was determined from temperature, luminosity, and stellar mass, where the luminosity was obtained by applying a bolometric correction of \citet{Casagrande2018a} to the absolute magnitude in the Gaia $G$ band, and mass was derived from isochrone fitting in the color-magnitude diagram. 
This method of $\log g$ determination has an advantage over the spectroscopic method, requiring the ionization equilibrium between Fe~I and Fe~II, in that it does not strongly depend on the assumed $T_{\rm eff}$.
For the stellar parameter determination, we used the simple mean of iron abundances derived from individual Fe~II lines because Fe II abundances are less affected by the non-LTE effect and temperature uncertainty than Fe I abundances. 
We took correlated uncertainties into consideration for the estimates of uncertainties.
The derived parameters are presented in Table~\ref{tab:parameters}.

The elemental abundances are the weighted average of the abundances derived from the individual lines.
The weights were determined following the prescription by \citet{Ji2020a}.
The weights on individual lines as well as line-by-line abundances and their sensitivities to stellar parameters are summarized in Table~\ref{tab:linelist}.
When considering [{X}/{Fe}], we used the Fe abundance from the same ionization stage as the species X.

All of the analysis was conducted relative to HD59392, for which we adopted $T_{\rm eff}=6012\,\mathrm{K}$, $\log g=3.954$, $v_t=1.4\,\mathrm{km\,s^{-1}}$, and $[\mathrm{Fe/H}]_{\rm sp}=-1.6$ (NS10; Paper I).
The elemental abundance of this star was taken from \citetalias{Nissen2010} and \citet{Nissen2011}.
In \citetalias{Matsuno2022a}, we validated our results using a few standard stars and confirm that our abundances are on the same scale as \citetalias{Nissen2010}, \citet{Nissen2011} and \citetalias{Reggiani2017a}.
We note that one of the standard stars was G112-43 in the Helmi streams, for which \citet{Nissen2011} and \citet{Nissen2021a} reported high Zn abundance. 
Since we do not find any systematic offset in the derived abundance for this star when comparing with \citetalias{Nissen2010} and \citet{Nissen2011}, our analysis also confirms its high Zn abundance.

In case no lines were detected for a species, we provided 1-$\sigma$ and 3-$\sigma$ upper limits on the abundance, which were estimated from the expected uncertainty on equivalent widths from the equation in \citet{Cayrel1988a}.
We used the Si~I line at $6237\,\mathrm{\AA}$ and the Y~II line at $4884\,\mathrm{\AA}$ to place upper limits.
The derived abundances and upper limits are presented in Table~\ref{tab:abundance}.

As an additional test, we investigated the effect of slightly different atmospheric structures due to different $\alpha$-element abundances in model atmospheres, since the sample includes stars showing different [$\alpha$/{Fe}] compared to typical halo stars. 
We used two MARCS model atmospheres with $T_{\rm eff}=6000\,\mathrm{K}$, $\log g =4.0$, $v_t=1.5\,\mathrm{km\,s^{-1}}$, and $[\mathrm{Fe/H}]=-1.5$ differing in $[\mathrm{\alpha/Fe}]$, one of which has the ``standard composition'' at this metallicity with $[\mathrm{\alpha/Fe}]=+0.4$ and the other has a solar $[\mathrm{\alpha/Fe}]$. 
Assuming the chemical composition of HD59392, we first computed equivalent widths of absorption lines using the standard $\alpha$-enhanced model atmosphere.
We then derived abundances with the other model from the computed equivalent widths.
The maximum difference between the assumed and derived abundances is found in strong lines to be $\sim 0.02\,\mathrm{dex}$.
Since the variation in $[\mathrm{\alpha/Fe}]$ among our sample is $\sim 0.2$ dex and since every line is affected in the same direction, the effect of different $[\mathrm{\alpha/Fe}]$ in model atmospheres is at most $0.01\,\mathrm{dex}$ when discussing [{X}/{H}] and even smaller when discussing [{X}/{Fe}].

\subsection{Results\label{sec:results}}
\begin{figure*}
\includegraphics[width=\textwidth]{\figureloc 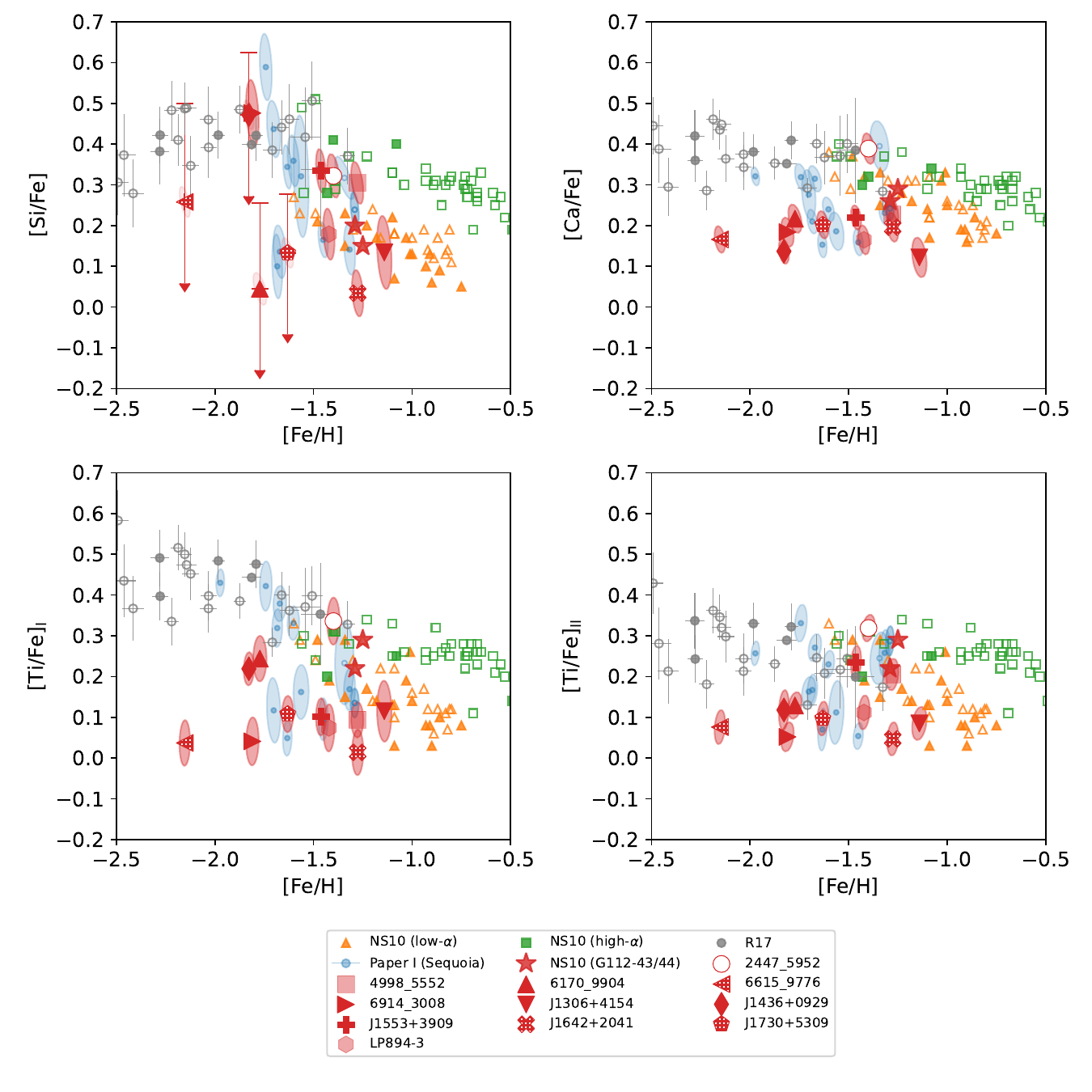}
\caption{$\alpha$-element abundances of the Helmi stream stars. The symbols follow those of Figure~\ref{fig:NaMg}. When a star does not have any detectable lines for an element, we indicate the abundances corresponding to the 1-$\sigma$ and 3-$\sigma$ upper limits on the equivalent widths, respectively shown with the location of the symbols and the upper end of the error bars. 
\label{fig:alpha}}
\end{figure*}

\begin{figure*}
\includegraphics[width=\textwidth]{\figureloc 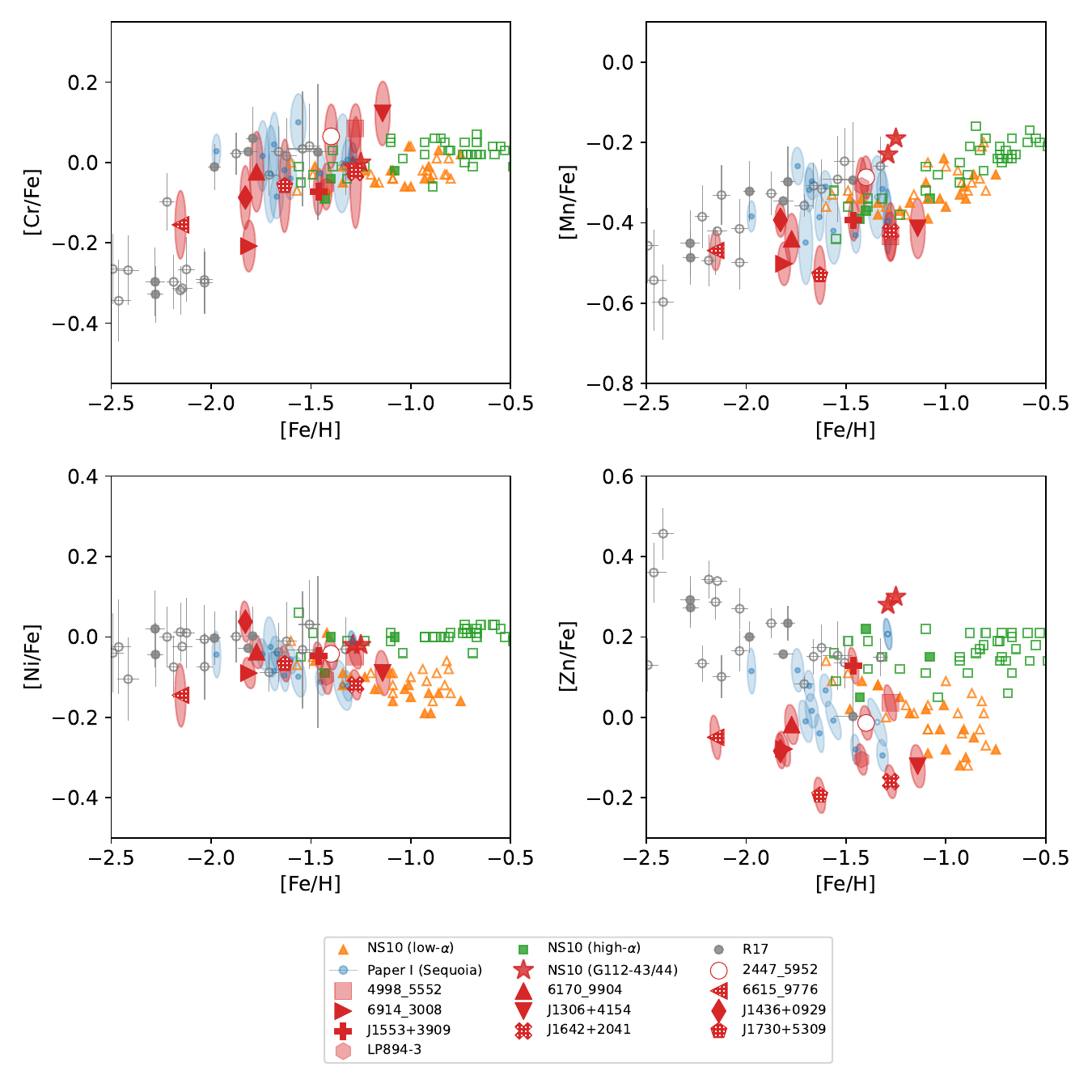}
\caption{Abundances of Cr, Mn, Ni, and Zn. Symbols follow Figure~\ref{fig:NaMg}.\label{fig:ironpeak}}
\end{figure*}

\begin{figure*}
\includegraphics[width=\textwidth]{\figureloc 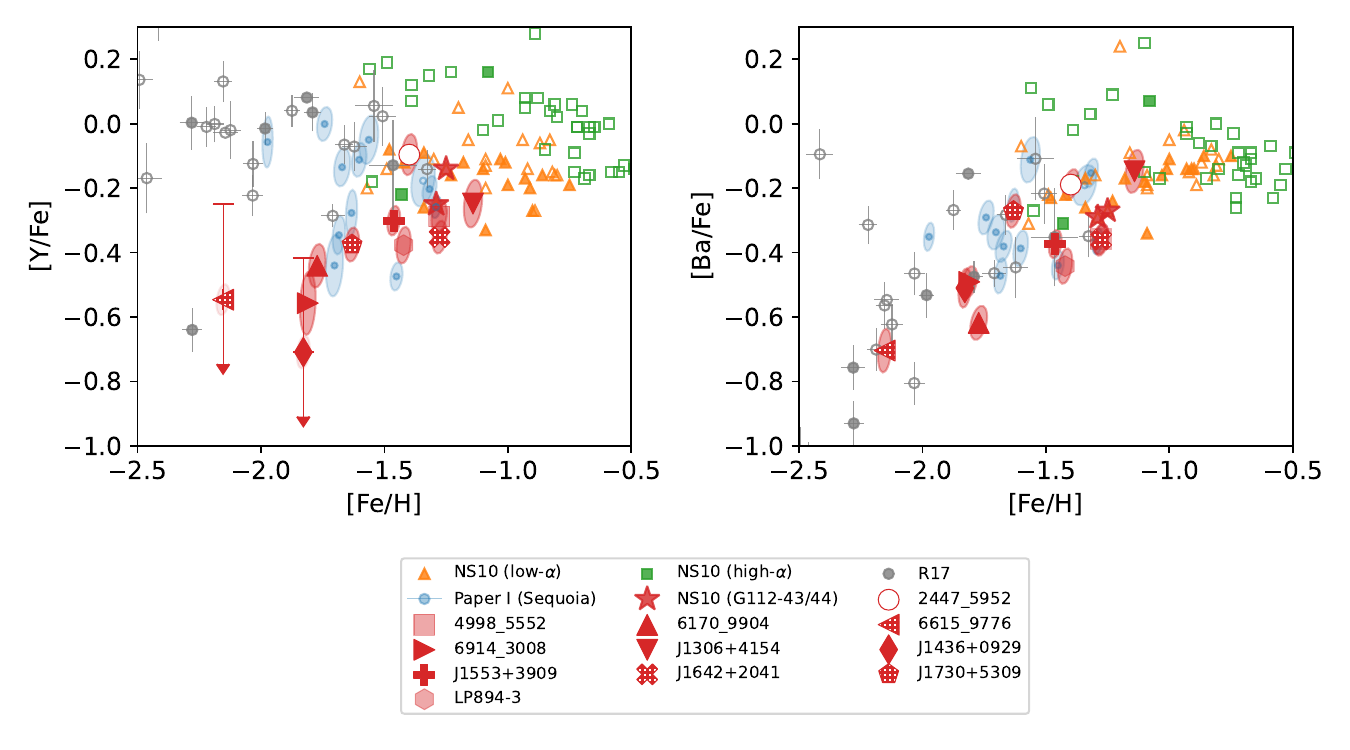}
\caption{Abundances of Y and Ba. Symbols follow Figure~\ref{fig:NaMg}.\label{fig:ncapture}}
\end{figure*}

Figure~\ref{fig:NaMg} shows [{Na}/{Fe}] and [{Mg}/{Fe}] ratios of the Helmi stream stars together with the halo stars from \citetalias{Nissen2010} and \citetalias{Reggiani2017a} and the Sequoia stars from \citetalias{Matsuno2022a}, all of which are on the same abundance scale.
We note that we indicate Gaia-Enceladus stars in the literature with filled symbols in figures.
At $-1.8\lesssim[\mathrm{Fe/H}]\lesssim -1.6$, the Helmi stream stars show similar Na and Mg abundances to Sequoia stars, having clearly lower [{Na}/{Fe}] and [{Mg}/{Fe}] values compared to other halo stars including Gaia-Enceladus stars.
The Helmi stream stars with $[\mathrm{Fe/H}]\gtrsim -1.5$ also have lower values in these abundance ratios than in situ halo stars.
Although they are less separated from Gaia-Enceladus stars at this high metallicity, they still tend to show lower [{Mg}/{Fe}] values compared to Gaia-Enceladus stars.
The most metal-poor star 6615\_9776 also seems to show a lower abundance of Na and Mg compared to other halo stars.
We, however, note that this star is in the metallicity range, $[\mathrm{Fe/H}]<-2.1$, where \citetalias{Reggiani2017a} measured the abundance of stars using a standard star that is different from what they used for stars with $[\mathrm{Fe/H}]>-2.1$.
Hence, there is no guarantee that the abundance of 6615\_9976 and \citetalias{Reggiani2017a} stars with similar metallicity are on the same scale. 
Readers can refer to \citetalias{Matsuno2022a} for the effect of different standard stars (see the results of abundance comparison for HIP28104 in Section 3 of Paper I).

Figure~\ref{fig:alpha} shows Si, Ca, and Ti abundances, which are also known to be different between stars with different origins (NS10; Paper I). 
The behavior of the Helmi stream stars in Ca and Ti is similar to that in Na and Mg; 
the stars show clearly lower abundance at low metallicity than most halo stars, while they become more indistinguishable from Gaia-Enceladus stars at higher metallicity.
For Si, the picture is unclear because the Si lines become too weak at $[\mathrm{Fe/H}]\lesssim -1.5$.

The observed trends in the $\alpha$-elements confirm the results of \citet{Limberg2021a} and \citet{Aguado2021b} in the sense that the Helmi stream stars have lower $\alpha$-element abundance compared to general Milky Way stars. 
We added here another important result; the $\alpha$-element abundances of the Helmi stream stars are even lower than another accreted population, the Gaia-Enceladus stars, especially at $[\mathrm{Fe/H}]\lesssim -1.5$.
Moreover, contrary to the conclusion of \citet{Limberg2021a} and \citet{Aguado2021b} that the $[\mathrm{\alpha/Fe}]$ ratios decrease with metallicity, the evolution in $\alpha$-elements (Mg, Ca, and Ti) with metallicity observed in the present study is rather flat from $[\mathrm{Fe/H}]\sim -2.1$ to $[\mathrm{Fe/H}]\sim -1.2$.
The low $\alpha$ abundance of the Helmi stream stars is also in contrast with \citet{Roederer2010a}, who concluded that the Helmi stream stars have a similar abundance as other halo stars.
We consider that our precise and homogeneous abundance comparisons enable us to detect the abundance difference between the Helmi stream stars and Gaia-Enceladus stars and precisely depict the chemical evolutionary track followed by the Helmi streams.
We further discuss the importance of the homogeneous abundance analysis in Section~\ref{sec:literature}.

The abundances of elements near the iron-peak, namely Cr, Mn, Ni, and Zn, are shown in Figure~\ref{fig:ironpeak}.
It has been reported that Cr, Ni and Zn abundances are different between accreted stars and in situ stars \citep{Nissen2011}.
We, therefore, expect that these abundances are different among the various kinematic substructures with different $\alpha$-element abundances.
We, however, do not see apparent abundance differences between the Helmi stream stars and the other halo stars in Cr and Ni at our measurement uncertainty.
On the other hand, Zn appears depleted in the Helmi streams at low metallicity, while there is a large variation at high metallicity.
The Helmi stream stars tend to show a low value of [{Mn}/{Fe}] at $[\mathrm{Fe/H}]\lesssim -1.5$ while they seem to be consistent with typical Gaia-Enceladus stars at higher metallicity.

The high Zn abundance of G112-43 and G112-44 was  reported in \citet{Nissen2011} and confirmed in \citet{Nissen2021a}.
It is also supported for G112-43 by our analysis in \citetalias{Matsuno2022a} as described in Section~\ref{sec:analysis}.
The other Helmi stream stars in the present study, on the other hand, do not share a similar Zn enhancement. 
We attempt to interpret the Zn abundance evolution of the Helmi streams in Section~\ref{sec:evolution}.

Abundances of neutron-capture elements are shown in Figure~\ref{fig:ncapture}.
The Helmi stream stars are clearly deficient in Y at low metallicity. 
This behavior of Y resembles that of Sr reported by \citet{Aguado2021b}, which is expected since both are light neutron-capture elements and since these two elements usually show similar trends with [{Fe}/{H}].
It is hard to conclude if the Ba abundance evolution of the Helmi streams is distinct from that of Gaia-Enceladus or other halo stars because of large scatter.
However, we remark that the Helmi stream stars tend to be on the lower side in [{Ba}/{Fe}] at given [{Fe}/{H}] compared to other halo stars.

As described in Section~\ref{sec:targets}, symbols in Figures~\ref{fig:NaMg} to \ref{fig:ncapture} are hatched according to the subgroups discussed by \citet{Dodd2022a}.
Both subgroups show similar abundance peculiarities compared to other halo stars (lower abundances of $\alpha$-elements and Y). 
This result supports that both subgroups can be regarded as a part of the Helmi streams.
We made a similar inspection between stars with positive and negative $v_z$ and reached the same conclusion. 

We note that the only star with an exceptionally high value of $E_n$ (2447\_5952) has a slightly different elemental abundance than the other Helmi stream stars.
The different behavior of this star is especially prominent in Ca, Ti and neutron-capture elements (Figures~\ref{fig:alpha} and \ref{fig:ncapture}). 
This might indicate that the Helmi streams do not extend toward high $E_n$, which is consistent with the recent identification of Helmi stream members based on clustering analysis of halo stars \citep{Loevdal2022a,RuizLara2022a}.

\section{Discussion\label{sec:discussion}}

\subsection{Properties of the Helmi stream progenitor \label{sec:evolution}}

The chemical properties of the Helmi streams can be summarized as follows.
The abundances of elements usually produced by explosions of massive stars (e.g., Na and $\alpha$-elements) are generally lower than most of the other halo stars, including Gaia-Enceladus stars.
While the difference to Gaia-Enceladus stars becomes less clear at $[\mathrm{Fe/H}]\gtrsim -1.5$, the Helmi stream stars still tend to be on the lower side of the distribution in [{Mg}/{Fe}], [{Ca}/{Fe}], and [{Ti}/{Fe}].
Zn and Y also show similar trends as these elements, suggesting they originate in massive stars.
This chemical abundance pattern is similar to that seen for Sequoia stars  \citepalias{Matsuno2022a}.

The low abundance of $\alpha$-elements and Na are usually attributed to Fe enrichments from type~Ia supernovae (SNe~Ia). 
Interestingly, SNe~Ia seem to have started to operate in the Helmi stream progenitor already at very low metallicity ($[\mathrm{Fe/H}]<-2.2$) since even the lowest-metallicity star in the sample (6615\_9776) has low abundances of $\alpha$-elements and Na. 
Assuming that every system has the same abundance ratios among Na, $\alpha$-elements, and Fe before the onset of SNe~Ia, 
we can tentatively provide an upper limit for the ``knee'' metallicity as $[\mathrm{Fe/H}]_{\rm knee} \lesssim -2.4$ from the $\sim 0.2\,\mathrm{dex}$ lower [{Mg}/{Fe}] of 6615\_9776 than the halo stars from \citetalias{Reggiani2017a}.

Another interesting feature in $\alpha$-elements is that there is no apparent decreasing trend in $[\mathrm{\alpha/Fe}]$ with metallicity.
Moreover, a few stars at $[\mathrm{Fe/H}]\sim -1.2$ seem to have higher $[\mathrm{\alpha/Fe}]$ than the stars at $[\mathrm{Fe/H}]\sim -1.8$, although it remains to be seen if they are actually the Helmi stream members rather than contaminants from heated thick disk or other accreted populations.
If this almost flat or slightly increasing trend is confirmed with a larger sample of homogeneous abundances, it would resemble the observed trend in the Large Magellanic Could (LMC) by \citet{Nidever2020a}.
These authors suggest that such a chemical evolution trend cannot be reproduced in models with a constant star formation efficiency.
They suggest that such a trend needs a starburst at a late phase.
Although the progenitor of the Helmi streams might have experienced a similar starburst at a late phase, the old age of the majority of the Helmi stream stars \citep{Koppelman2019b,RuizLara2022a} suggests that the transition from quiescent to bursty star formation occurred on a shorter timescale than that in LMC.
A future chemodynamical study with precise stellar age as well as chemical evolution modeling would be necessary to confirm this scenario.

The Y abundance of the Helmi stream stars is particularly low at $[\mathrm{Fe/H}]\lesssim -1.8$.
Together with the results of \citet{Aguado2021b} on Sr abundance, this suggests that the Helmi stream stars have very low abundances of light neutron-capture elements. 
Such a low light neutron-capture element abundance is not common among the Milky Way stars but is seen in low-mass dwarf galaxies such as Draco and Ursa Minor dwarf spheroids and most of the ultra-faint dwarf galaxies  \citep[e.g.,][]{Frebel2015}. 

Several nucleosynthesis processes have been proposed as production sites of these light neutron-capture elements, including $r$-process in neutron star mergers \citep{Wanajo2014a,Watson2019a}, in magneto-rotational supernovae \citep{Winteler2012a}, or in collapsars \citep{Siegel2019a}, $s$-process in low- to intermediate-mass stars \citep{Karakas2014}, weak $s$-process in rapidly rotating massive stars \citep{Frischknecht2012a,Choplin2018a}, and weak $r$-process in electron-capture supernovae \citep{Wanajo2011a}.
It is not yet clear which process is the dominant source of the elements in the early Universe \citep[see discussions by, e.g., ][]{Cote2019a,Prantzos2018a,Kobayashi2020a}.
Since we here discuss the low-metallicity end of the sample, the production of neutron-capture elements would not be dominated by low- to intermediate-mass stars \citep[e.g.,][]{Reyes2022a}.
One possible explanation for the low light neutron-capture element abundances of the Helmi streams is that, as a result of the low stellar mass of the galaxy, the progenitor did not experience rare $r$-process nucleosynthesis events, such as neutron star mergers, electron capture supernovae, and magneto-rotational supernovae. 
In this case, a small amount of light neutron-capture elements could be produced by rapidly rotating massive stars \citep{Hirai2019a,Tarumi2021a}.
Another explanation is that the progenitor dwarf galaxy had a small number of rotating massive stars.
The small number of rotating massive stars might be a result of the top-light initial mass function in dwarf galaxies \citep{Weidner2005a}, or different distribution of initial rotation velocity of stars.
The observational indication by \citet{Gull2021a} that metal-poor stars of the Helmi streams show $r$-process abundance pattern in neutron-capture elements heavier than Ba might favor the second possibility.
However, it is necessary to investigate the abundance pattern of light neutron-capture elements in order to understand the cause of the low light neutron-capture element abundance of the Helmi streams. 
A larger sample of low-metallicity Helmi stream stars with neutron-capture element abundances would also be welcomed.
They would enable us to constrain the property of the nucleosynthesis processes, such as their event rates, by studying how neutron-capture elements were enriched as a function of metallicity \citep[e.g.,][]{Tsujimoto2017a}.

The [{Zn}/{Fe}] values observed at high metallicity are also noteworthy.
While the binary pair G112-43 and G112-44 have high Zn abundance \citep{Nissen2021a}, other Helmi stream stars do not share such a high abundance, resulting in a large star-to-star variation in [{Zn}/{Fe}].
Although \citet{Nissen2021a} noted that the binary pair is also enhanced in Mn, Ni, and Cu, the measurement uncertainties in the present study are not high enough to see if there are significant scatters in Mn and Ni abundances among the Helmi stream stars.
A significant dispersion in [{Zn}/{Fe}] at high metallicity is also reported in the Sculptor dwarf spheroidal galaxy \citep{Skuladottir2017}.
While \citet{Hirai2018a} provided a theoretical calculation of Zn enrichments in dwarf galaxies, assuming electron capture supernovae as one of the sources of Zn, they find it challenging to produce a large scatter at high metallicity.
Based on iron-group element abundances, \citet{Nissen2021a} suggest pure helium detonation type~Ia supernovae could be a promising nucleosynthesis event producing the unique abundance pattern of the binary pair.
It remains to be seen whether chemical evolution models including this type of SNe~Ia can explain the Zn abundance variation within a galaxy. 
A larger sample of stars with precise multi-element abundances would be helpful to establish, for example, if the binary pair is a chemical outlier or a tail of [{Zn}/{Fe}] distribution in the system and if the Zn abundance correlates with abundances of other elements.

\subsection{Helmi streams in the literature\label{sec:literature}}

\begin{figure*}
\includegraphics[width=\textwidth]{\figureloc 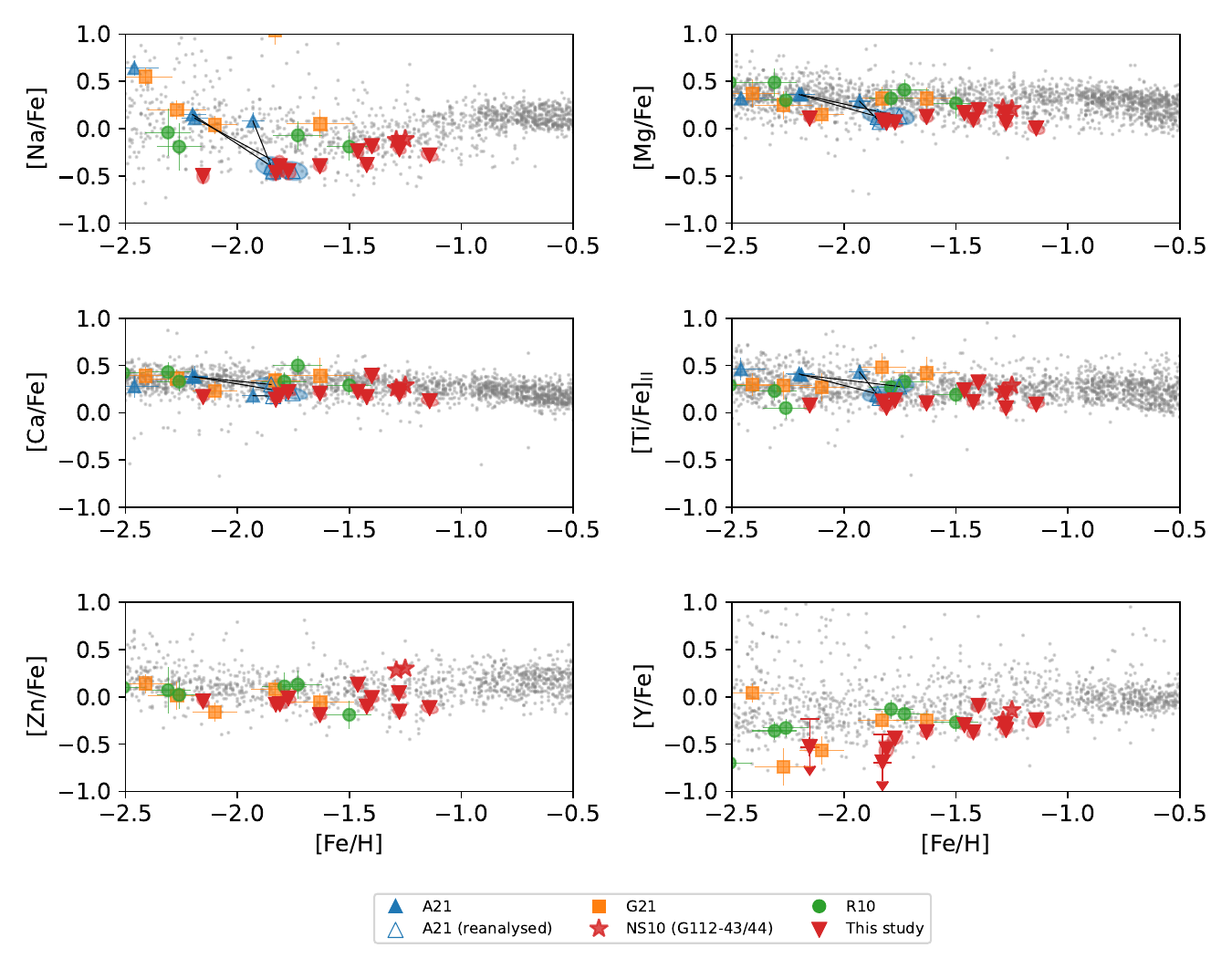}
\caption{Chemical abundance of the Helmi stream stars from literature, namely \citet[R10; ][]{Roederer2010a}, \citet[A21; ][]{Aguado2021b}, and \citet[G21; ][]{Gull2021a}. We also include Helmi stream stars in the present study and in \citetalias{Nissen2010}. The background gray points are from the SAGA database \citep{Suda2008,Suda2011,Yamada2013}. The UVES spectra of three stars from \citet{Aguado2021b} were reanalysed and abundances from the reanalysis are included in the figure. Abundances from our reanalysis and those from \citet{Aguado2021b} are connected with a solid line for each star. \label{fig:lite_comparison}}
\end{figure*}

In this section, we highlight the importance of homogeneous chemical abundance through a comparison of our results with those presented in the literature, by \citet{Roederer2010a}, by \citet{Limberg2021a}, by \citet{Gull2021a}, and by \citet{Aguado2021b}\footnote{We consider five stars observed with high-resolution spectrographs (HORuS and UVES) for the \citet{Aguado2021b} sample.}. 
All but \citet{Limberg2021a} conducted new observations of the Helmi stream stars, while \citet{Limberg2021a} used the data from GALAH DR3.
All, including \citet{Limberg2021a}, compared the chemical abundance of the Helmi stream stars with a literature compilation without homogenizing the chemical abundances.

Figure~\ref{fig:lite_comparison} compares the abundances of the Helmi stream stars from those literature studies and from our present work, with those of stars in the Milky Way from various literature sources \citep[SAGA database; ][]{Suda2008,Suda2011,Yamada2013}.
To confirm the membership to the Helmi streams, we recomputed angular momenta of the stars in \citet{Roederer2010a}, \citet{Gull2021a}, and \citet{Aguado2021b} consistently and reselected the Helmi stream members.
As there is no report of radial velocity in \citet{Aguado2021b}, we adopted radial velocity measurements by SDSS or LAMOST for their stars.
For the stars in common between \citet{Roederer2010a} and \citet{Gull2021a}, we prioritized the measurements by the latter study over the former.
We also included in the figure abundances that we obtained for three stars studied by \citet{Aguado2021b} by reanalyzing their UVES spectra.
For this reanalysis, we redetermined stellar parameters following the same procedure as described in \citetalias{Matsuno2022a} and estimated the abundances using these parameters.

We first emphasize the importance of homogeneity in abundance by using the three stars from \citet{Aguado2021b} that we reanalyzed.
As can be seen in Figure~\ref{fig:lite_comparison}, stars from \citet{Aguado2021b} seem to have higher $\alpha$-elements abundances compared to the Helmi stream stars in the present study if we adopt the reported abundances.
On the other hand, our reanalysis shows that the three stars actually have similar abundances as the sample in the present study, although the uncertainties are larger due to different spectral quality.
To investigate the effect of different stellar parameters, we fixed $T_{\rm eff}$ and $\log g$ to the values adopted by \citet{Aguado2021b}, determined $v_t$ and metallicity, and then measured abundances instead of redetermining the four stellar parameters.
The results do not change significantly compared to what we obtained from the full reanalysis.
The source of difference thus could be due to different atomic data, line selection, and/or spectral synthesis software.

The different abundance from \citet{Aguado2021b} and our reanalysis is not too surprising since it has been known that systematic uncertainty can be significant when abundances from different studies are compared without homogenization.
For example, G112-43 is a well studied star and its abundance is available in a number of literature \citep{Ryan2001a,Charbonnel2005,Zhang2009a,Ishigaki2010,Nissen2010,Nissen2011,Ishigaki2012,Ishigaki2013,Yan2016a}\footnote{The data are collected through the SAGA database.}.
The reported metallicity ($[\mathrm{Fe/H}]_{\mathrm{I}}$) and [{Mg}/{Fe}] of this star respectively range from $-1.52$ \citep{Ishigaki2010} to $-1.14$ \citep{Zhang2009a}, and from 0.16 \citep{Zhang2009a} to 0.37 \citep{Ishigaki2010}.
These ranges are significantly larger than the typical uncertainty reported.
We note that we derive $[\mathrm{Fe/H}]_{\mathrm{I}}=-1.254$ and $[\mathrm{Mg/Fe}]=0.179$ in \citetalias{Matsuno2022a} and \citetalias{Nissen2010} (adopted for figures) derive $[\mathrm{Fe/H}]_{\mathrm{I}}=-1.25$ and $[\mathrm{Mg/Fe}]=0.21$.

Figure~\ref{fig:lite_comparison} demonstrates that our sample has the smallest uncertainty, which was made possible thanks to the high quality of the spectra.
However, even with the high precision, the chemical distinctness of the Helmi stream stars is less clear than what we see in Figures~\ref{fig:NaMg}--\ref{fig:ncapture}. 
This is because the comparison sample from the SAGA database is not necessarily on the same abundance scale as ours, because its distribution is broadened due to systematic uncertainty, and because it can contain poor quality data.

In conclusion, systematic uncertainties have hampered a clear chemical characterization of the Helmi stream stars in previous studies.
Thanks to our approach of abundance analysis and homogenization of abundances from literature \citepalias[see ][]{Matsuno2022a}, we can detect clear but small abundance difference between the Helmi stream stars and other halo stars that include Gaia-Enceladus stars.

\section{Summary\label{sec:conclusion}}

Through a differential abundance analysis, we show that the Helmi stream stars clearly depict lower [{X}/{Fe}] in Na, Mg, Ca, Ti, Zn, and Y, compared to the other halo stars, including stars accreted from Gaia-Enceladus.
While the distinction from Gaia-Enceladus stars is clearer at $[\mathrm{Fe/H}]\lesssim -1.5$, a part of the Helmi stream stars start to overlap with Gaia-Enceladus stars in elemental abundance ratios at high metallicities.

This is the first time that we see the chemical distinctness of the Helmi stream stars over a wide metallicity range.
This result was made possible thanks to the precise abundances derived in this study and to the homogenization of abundances across \citetalias{Nissen2010}, \citetalias{Reggiani2017a} and the present study.
As in \citetalias{Matsuno2022a}, the present investigation of the Helmi streams highlights the importance of homogenized abundances for chemical identification of individual building blocks of the Milky Way. 

The observed low $\alpha$-element abundances of the Helmi streams are likely due to the Fe enrichments from SNe~Ia at low metallicity.
There seems to be a contribution of SNe~Ia to the chemical evolution of the progenitor already at very low metallicity ($[\mathrm{Fe/H}]\lesssim -2.4$), indicating that the star formation proceeded with low efficiency in early times.
Star formation at later times might have been bursty since [$\alpha$/{Fe}] remains almost constant or slightly increases toward high metallicities according to \citet{Nidever2020a};
such a feature may require a late starburst.

The origin of the extremely low Y abundances at low metallicity remains unclear.
Our results, as well as those of \citet{Aguado2021b} on Sr, strongly suggest that the Helmi streams have very low abundances of light neutron-capture elements at low metallicity.
It would be interesting to obtain detailed abundance patterns over light neutron-capture elements in order to constrain the nucleosynthesis processes that produce this feature.

Further observations would also be able to aid understanding of the Zn abundance variation at high metallicity.
The binary pair G112-43 and G112-44 are the only stars with clear Zn enhancements, and all the other Helmi stream stars at $[\mathrm{Fe/H}]\gtrsim -1.5$ have Zn abundances comparable to or lower than Gaia-Enceladus stars.
It would be interesting to investigate the detailed [{Zn}/{Fe}] distribution with a lager number of stars.

\begin{acknowledgements}
We thank the anonymous referee for his/her constructive comments, which helped us with improving the clarity of the discussion.
We thank David Aguado for sharing their UVES spectra with us.
This research has been supported by a Spinoza Grant from the Dutch Research Council (NWO).
WA and MNI were supported by JSPS KAKENHI Grant Number 21H04499. MNI is supported by JSPS KAKENHI Grant Number 20H05855.
ZY acknowledges support from the French National Research Agency (ANR) funded project ``Pristine'' (ANR-18-CE31-0017) and the European Research Council (ERC) under the European Unions Horizon 2020 research and innovation programme (grant agreement No. 834148).
This research is based in part on data collected at Subaru Telescope, which is operated by the National Astronomical Observatory of Japan.
We are honored and grateful for the opportunity of observing the Universe from Maunakea, which has the cultural, historical and natural significance in Hawaii.
Part of the data were retrieved from the JVO portal (http://jvo.nao.ac.jp/portal/) operated by ADC/NAOJ.
\end{acknowledgements}


\begin{appendix}
\section{Additional table}

\longtab[1]{
\begin{landscape}
\begin{longtable}{l*{17}{r}}
\caption{Abundances of the Helmi stream stars\label{tab:abundance}}\\
\hline\hline
\endfirsthead
\caption{continued.}\\
\hline\hline
\endhead
\endfoot
     &                     \multicolumn{5}{c}{2447\_5952}&     &                     \multicolumn{5}{c}{4998\_5552}&&                     \multicolumn{5}{c}{6170\_9904}\\\cline{2-6}\cline{8-12}\cline{14-18}
     & $N$ & [{X}/{H}] & $\sigma$ & [{X}/{Fe}] & $\sigma$&     & $N$ & [{X}/{H}] & $\sigma$ & [{X}/{Fe}] & $\sigma$&& $N$ & [{X}/{H}] & $\sigma$ & [{X}/{Fe}] & $\sigma$\\\hline
FeI  &    103&    -1.399&     0.032&       ...&       ...&&     88&    -1.278&     0.031&       ...&       ...&     &     93&    -1.772&     0.034&       ...&       ...\\
FeII &     12&    -1.420&     0.025&       ...&       ...&&     13&    -1.268&     0.042&       ...&       ...&     &     11&    -1.733&     0.025&       ...&       ...\\
NaI  &      3&    -1.585&     0.055&    -0.186&     0.054&&      2&    -1.430&     0.041&    -0.152&     0.039&     &      2&    -2.222&     0.093&    -0.450&     0.089\\
MgI  &      7&    -1.202&     0.041&     0.196&     0.042&&      4&    -1.119&     0.040&     0.159&     0.038&     &      5&    -1.706&     0.052&     0.066&     0.050\\
SiI  &      2&    -1.077&     0.051&     0.322&     0.054&&      6&    -0.973&     0.046&     0.305&     0.052&     &      0&\multicolumn{2}{c}{$<$-1.727(-1.517)} &  \multicolumn{2}{c}{$<$0.045(0.255)}\\
CaI  &     19&    -1.010&     0.041&     0.389&     0.038&&     18&    -1.051&     0.044&     0.227&     0.040&     &     19&    -1.555&     0.038&     0.217&     0.035\\
TiI  &     11&    -1.063&     0.067&     0.336&     0.058&&     15&    -1.184&     0.053&     0.094&     0.044&     &      8&    -1.528&     0.066&     0.244&     0.056\\
TiII &     13&    -1.102&     0.041&     0.319&     0.033&&      8&    -1.064&     0.043&     0.205&     0.039&     &     12&    -1.604&     0.035&     0.129&     0.033\\
CrI  &      4&    -1.334&     0.087&     0.065&     0.080&&      2&    -1.193&     0.070&     0.085&     0.061&     &      3&    -1.796&     0.106&    -0.023&     0.100\\
MnI  &      5&    -1.686&     0.060&    -0.287&     0.053&&      7&    -1.712&     0.068&    -0.434&     0.062&     &      5&    -2.212&     0.071&    -0.440&     0.063\\
NiI  &     14&    -1.441&     0.045&    -0.042&     0.040&&     22&    -1.327&     0.031&    -0.049&     0.031&     &      8&    -1.810&     0.047&    -0.037&     0.042\\
ZnI  &      2&    -1.412&     0.047&    -0.014&     0.043&&      2&    -1.242&     0.037&     0.036&     0.045&     &      2&    -1.790&     0.052&    -0.018&     0.050\\
YII  &      2&    -1.516&     0.069&    -0.096&     0.064&&      2&    -1.556&     0.047&    -0.288&     0.049&     &      1&    -2.174&     0.069&    -0.441&     0.068\\
BaII &      4&    -1.610&     0.058&    -0.189&     0.048&&      4&    -1.627&     0.049&    -0.358&     0.048&     &      4&    -2.351&     0.055&    -0.618&     0.053\\\hline
     &                     \multicolumn{5}{c}{6615\_9776}&&                     \multicolumn{5}{c}{6914\_3008}&&                     \multicolumn{5}{c}{J1306+4154}\\\cline{2-6}\cline{8-12}\cline{14-18}
     & $N$ & [{X}/{H}] & $\sigma$ & [{X}/{Fe}] & $\sigma$&& $N$ & [{X}/{H}] & $\sigma$ & [{X}/{Fe}] & $\sigma$&& $N$ & [{X}/{H}] & $\sigma$ & [{X}/{Fe}] & $\sigma$\\\hline
FeI  &     92&    -2.153&     0.027&       ...&       ...&&     95&    -1.810&     0.032&       ...&       ...&&    102&    -1.141&     0.037&       ...&       ...\\
FeII &      5&    -2.183&     0.030&       ...&       ...&&     11&    -1.737&     0.027&       ...&       ...&&     13&    -1.214&     0.030&       ...&       ...\\
NaI  &      2&    -2.654&     0.083&    -0.501&     0.078&&      3&    -2.208&     0.116&    -0.398&     0.114&&      3&    -1.424&     0.059&    -0.283&     0.059\\
MgI  &      5&    -2.048&     0.046&     0.106&     0.044&&      7&    -1.750&     0.042&     0.061&     0.042&&      7&    -1.140&     0.059&     0.002&     0.059\\
SiI  &      0& \multicolumn{2}{c}{$<$-1.894(-1.653)} &  \multicolumn{2}{c}{$<$0.258(0.499)} &&      1&    -1.334&     0.079&     0.476&     0.082&&      3&    -1.007&     0.087&     0.134&     0.090\\
CaI  &     18&    -1.988&     0.034&     0.166&     0.033&&     19&    -1.626&     0.039&     0.184&     0.036&&     17&    -1.020&     0.051&     0.122&     0.049\\
TiI  &      8&    -2.117&     0.064&     0.037&     0.056&&     10&    -1.769&     0.069&     0.041&     0.059&&     11&    -1.027&     0.082&     0.115&     0.075\\
TiII &     10&    -2.107&     0.034&     0.076&     0.042&&      9&    -1.684&     0.029&     0.052&     0.036&&     11&    -1.129&     0.042&     0.085&     0.041\\
CrI  &      2&    -2.308&     0.090&    -0.155&     0.085&&      3&    -2.018&     0.073&    -0.208&     0.064&&      5&    -1.020&     0.089&     0.122&     0.080\\
MnI  &      2&    -2.623&     0.056&    -0.469&     0.050&&      4&    -2.313&     0.062&    -0.502&     0.054&&      6&    -1.556&     0.083&    -0.414&     0.075\\
NiI  &      5&    -2.298&     0.081&    -0.145&     0.079&&     12&    -1.900&     0.042&    -0.090&     0.039&&     13&    -1.231&     0.060&    -0.090&     0.056\\
ZnI  &      2&    -2.203&     0.052&    -0.050&     0.055&&      2&    -1.889&     0.040&    -0.079&     0.044&&      2&    -1.263&     0.057&    -0.122&     0.054\\
YII  &      0&  \multicolumn{2}{c}{$<$-2.729(-2.433)} &  \multicolumn{2}{c}{$<$-0.546(-0.250)}&&      1&    -2.294&     0.095&    -0.557&     0.098&&      2&    -1.462&     0.074&    -0.248&     0.074\\
BaII &      2&    -2.887&     0.062&    -0.704&     0.068&&      4&    -2.228&     0.043&    -0.491&     0.049&&      4&    -1.362&     0.069&    -0.148&     0.067\\\hline
     &                     \multicolumn{5}{c}{J1436+0929}&&                     \multicolumn{5}{c}{J1553+3909}&&                     \multicolumn{5}{c}{J1642+2041}\\\cline{2-6}\cline{8-12}\cline{14-18}
     & $N$ & [{X}/{H}] & $\sigma$ & [{X}/{Fe}] & $\sigma$&& $N$ & [{X}/{H}] & $\sigma$ & [{X}/{Fe}] & $\sigma$&& $N$ & [{X}/{H}] & $\sigma$ & [{X}/{Fe}] & $\sigma$\\\hline
FeI  &     88&    -1.828&     0.026&       ...&       ...&&     74&    -1.462&     0.026&       ...&       ...&&    103&    -1.276&     0.029&       ...&       ...\\
FeII &      8&    -1.822&     0.038&       ...&       ...&&      9&    -1.410&     0.027&       ...&       ...&&     15&    -1.249&     0.044&       ...&       ...\\
NaI  &      2&    -2.296&     0.082&    -0.468&     0.077&&      4&    -1.700&     0.068&    -0.238&     0.066&&      3&    -1.495&     0.049&    -0.219&     0.048\\
MgI  &      6&    -1.742&     0.047&     0.086&     0.044&&      4&    -1.298&     0.041&     0.164&     0.039&&      6&    -1.216&     0.045&     0.060&     0.044\\
SiI  &      0&  \multicolumn{2}{c}{$<$-1.355(-1.203)} &  \multicolumn{2}{c}{$<$0.472(0.624)} &&      2&    -1.128&     0.051&     0.334&     0.054&&      3&    -1.242&     0.056&     0.034&     0.057\\
CaI  &     13&    -1.691&     0.034&     0.137&     0.031&&     12&    -1.241&     0.033&     0.220&     0.030&&     16&    -1.080&     0.038&     0.196&     0.036\\
TiI  &      7&    -1.609&     0.050&     0.219&     0.041&&     14&    -1.361&     0.054&     0.101&     0.046&&     10&    -1.262&     0.064&     0.014&     0.056\\
TiII &     12&    -1.704&     0.039&     0.118&     0.049&&      2&    -1.175&     0.039&     0.235&     0.041&&     10&    -1.202&     0.051&     0.047&     0.041\\
CrI  &      2&    -1.915&     0.084&    -0.087&     0.079&&      3&    -1.535&     0.065&    -0.073&     0.058&&      2&    -1.301&     0.142&    -0.025&     0.139\\
MnI  &      3&    -2.221&     0.049&    -0.393&     0.043&&      4&    -1.856&     0.056&    -0.394&     0.051&&      5&    -1.698&     0.077&    -0.422&     0.072\\
NiI  &      5&    -1.790&     0.053&     0.038&     0.051&&     18&    -1.509&     0.038&    -0.047&     0.035&&     12&    -1.395&     0.041&    -0.119&     0.038\\
ZnI  &      2&    -1.912&     0.045&    -0.084&     0.044&&      2&    -1.333&     0.041&     0.128&     0.046&&      2&    -1.436&     0.042&    -0.160&     0.042\\
YII  &      0&  \multicolumn{2}{c}{$<$-2.531(-2.238)} &  \multicolumn{2}{c}{$<$-0.709(-0.416)} &&      2&    -1.711&     0.042&    -0.301&     0.046&&      2&    -1.601&     0.056&    -0.352&     0.051\\
BaII &      3&    -2.332&     0.052&    -0.510&     0.061&&      4&    -1.783&     0.043&    -0.373&     0.044&&      4&    -1.609&     0.057&    -0.360&     0.050\\\hline
     &                     \multicolumn{5}{c}{J1730+5309}&&                     \multicolumn{5}{c}{LP894-3   }\\\cline{2-6}\cline{8-12}
     & $N$ & [{X}/{H}] & $\sigma$ & [{X}/{Fe}] & $\sigma$&& $N$ & [{X}/{H}] & $\sigma$ & [{X}/{Fe}] & $\sigma$\\\cline{1-12}
FeI  &     84&    -1.631&     0.029&       ...&       ...&&     98&    -1.422&     0.029&       ...&       ...\\
FeII &     12&    -1.643&     0.039&       ...&       ...&&     14&    -1.488&     0.033&       ...&       ...\\
NaI  &      3&    -2.029&     0.062&    -0.398&     0.059&&      3&    -1.807&     0.043&    -0.385&     0.043\\
MgI  &      6&    -1.508&     0.057&     0.123&     0.054&&      6&    -1.329&     0.055&     0.093&     0.052\\
SiI  &      0&   \multicolumn{2}{c}{$<$-1.497(-1.352)} &  \multicolumn{2}{c}{$<$0.133(0.278)}   &&      4&    -1.243&     0.062&     0.179&     0.063\\
CaI  &     16&    -1.429&     0.037&     0.202&     0.034&&     18&    -1.257&     0.039&     0.165&     0.036\\
TiI  &      8&    -1.523&     0.055&     0.108&     0.045&&     10&    -1.348&     0.066&     0.074&     0.058\\
TiII &      9&    -1.546&     0.049&     0.097&     0.042&&      7&    -1.376&     0.037&     0.112&     0.042\\
CrI  &      2&    -1.689&     0.121&    -0.058&     0.116&&      4&    -1.482&     0.066&    -0.060&     0.058\\
MnI  &      7&    -2.161&     0.078&    -0.530&     0.073&&      3&    -1.730&     0.074&    -0.308&     0.070\\
NiI  &      9&    -1.698&     0.047&    -0.067&     0.045&&     16&    -1.522&     0.046&    -0.100&     0.043\\
ZnI  &      2&    -1.825&     0.045&    -0.194&     0.045&&      2&    -1.527&     0.040&    -0.105&     0.039\\
YII  &      2&    -2.018&     0.057&    -0.375&     0.053&&      2&    -1.865&     0.055&    -0.377&     0.057\\
BaII &      3&    -1.914&     0.057&    -0.271&     0.052&&      4&    -1.929&     0.054&    -0.442&     0.055\\\cline{1-12}
\end{longtable}
\tablefoot{For upper limits, we provide both 1-$\sigma$ and 3-$\sigma$ upper limits. The values in parenthesis are for the 3-$\sigma$ upper limits.
}
\end{landscape}
}

\end{appendix}
\end{document}